\documentclass[conference, 10pt]{IEEEtran}
\usepackage[dvips]{graphicx}
\usepackage{amssymb,amsmath}
\usepackage{subfig}
\usepackage{colortbl}
\usepackage{url}
\usepackage{cite}

\makeatletter
\def\ps@headings{%
\def\@oddhead{\mbox{}\scriptsize\rightmark \hfil \thepage}%
\def\@evenhead{\scriptsize\thepage \hfil \leftmark\mbox{}}%
\def\@oddfoot{}%
\def\@evenfoot{}}
\makeatother
\IEEEoverridecommandlockouts

\begin{document}

\title{Trade-off between cost and goodput in wireless:\\Replacing transmitters with coding
}
\author{
\authorblockN{MinJi Kim\authorrefmark{1}, Thierry Klein\authorrefmark{2}, Emina Soljanin\authorrefmark{2}, Jo\~{a}o Barros\authorrefmark{4}, Muriel M\'{e}dard\authorrefmark{1}\vspace*{-0.8cm}}
\thanks{\authorrefmark{1}M. Kim and M. M\'edard are with the RLE at the Massachusetts Institute of Technology, MA USA (e-mail: \{minjikim, medard\}@mit.edu).}
\thanks{\authorrefmark{2}T. Klein and E. Soljanin are with the Alcatel-Lucent Bell Laboratories, NJ USA (e-mail: thierry.klein@alcatel-lucent.com, emina@research.bell-labs.com).}
\thanks{\authorrefmark{4}J. Barros is with the Department of Electrical and Computer Engineering at the University of Porto, Portugal (e-mail: jbarros@fe.up.pt).}
}

\maketitle

\begin{abstract}
We study the cost of improving the \emph{goodput}, or the useful data rate, to user in a wireless network. We measure the cost in terms of number of base stations, which is highly correlated to the energy cost as well as capital and operational costs of a network provider. We show that increasing the available bandwidth, or throughput, may not necessarily lead to increase in goodput, particularly in lossy wireless networks in which TCP does not perform well. As a result, much of the resources dedicated to the user may not translate to high goodput, resulting in an inefficient use of the network resources. We show that using protocols such as TCP/NC, which are more resilient to erasures and failures in the network, may
lead to a goodput commensurate the throughput dedicated to each user. By increasing goodput, users' transactions are completed faster; thus, the resources dedicated to these users can be released to serve other requests or transactions. Consequently, we show that translating efficiently throughput to goodput may bring forth better connection to users while reducing the cost for the network providers.

\end{abstract}
\IEEEpeerreviewmaketitle

\section{Introduction}\label{sec:introduction}

\begin{figure}[tbp]
\begin{center}\vspace*{.4cm}
\includegraphics[width=.35\textwidth]{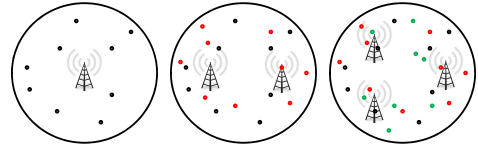}
\end{center}\vspace*{-.3cm}\caption{As number of users in a given area grows, a service provider may add additional base stations not for coverage but for bandwidth. As red users join the network, a second base station may be necessary; as green users join the network, a third base station may become necessary in order to maintain a certain level of quality of service.
}\label{fig:towers}\vspace*{-.4cm}
\end{figure}

Mobile data traffic has been growing at an alarming rate with some estimating that it will increase more than 25-folds in the next five years \cite{cisco}. In order to meet such growth, there has been an increasing effort to install and upgrade the current networks. As shown in Figure \ref{fig:towers}, mobile service providers often install more infrastructure (e.g. more base stations) in areas which already have full coverage. The new infrastructure is to provide more bandwidth, which would lead to higher quality of experience to users. However, this increase in bandwidth comes at a significant energy cost as each base station has been shown to use 2-3 kilowatts (kW) \cite{bell_power}. The sustainability and the feasibility of such rapid development have been brought to question as several trends indicate that the technology efficiency improvements may not be able to keep pace with the traffic growth \cite{bell_power}.

We show that maintaining or even improving users' quality of experience may be achieved without installing more base stations. In some cases, we show that the users' quality of experience may be improved while reducing the number of base stations. We measure users' quality of experience using the throughput perceived by the user or the application, i.e. \emph{goodput}. We make a clear distinction between the terms goodput and \emph{throughput}, where goodput is the number of \emph{useful} bits over unit time received by the user and throughput is the number of bits transmitted by the base station per unit time. In essence, throughput is indicative of the bandwidth/resources provisioned by the service providers; while goodput is indicative of the user's quality of experience. For example, the base station, taking into account the error correction codes, may be transmitting bits at 10 megabits per second (Mbps), i.e. throughput is 10 Mbps. However, the user may only receive useful information at 5 Mbps, i.e. goodput is 5 Mbps.

There can be a significant disparity between throughput and goodput, particularly in lossy networks using TCP. TCP often mistakes random erasures as congestion \cite{TCP_Kurose,hari}. For example, 1-3\% packet loss rate is sufficient to harm TCP's performance \cite{TCP_Kurose, hari, analysis, caceres}. This performance degradation can lead to inefficient use of network resources and incur substantially higher cost to maintain the same goodput. There has been extensive research to combat these harmful effects of erasures and failures; however, TCP even with modifications does not achieve significant improvement. References \cite{hari,Tian05tcpin} give an overview of various TCP versions over wireless links.

This disparity between throughput and goodput can be reduced by using a transport protocol that is more resilient to losses. One method is to use multiple base stations simultaneously (using multiple TCP connections \cite{parallel} or multipath TPC \cite{mptcp2}). However, the management of the multiple streams or paths may be difficult, especially in lossy networks. Furthermore, each path or TCP stream still suffer from performance degradation in lossy environments \cite{parallel, mptcp2}.

\begin{figure}[tbp]
\begin{center}\vspace*{.4cm}
\includegraphics[width=.35\textwidth]{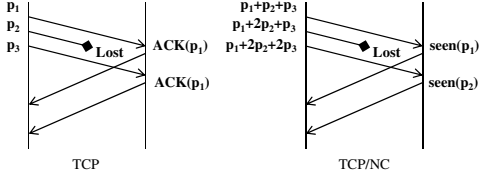}
\end{center}\vspace*{-.3cm}\caption{Example of TCP and TCP/NC. In the case of TCP, the TCP sender receives duplicate ACKs for packet $\mathbf{p_1}$, which may wrongly indicate congestion. However, for TCP/NC, the TCP sender receives ACKs for packets $\mathbf{p_1}$ and $\mathbf{p_2}$; thus, the TCP sender perceives a longer RTT but does not mistake the loss to be congestion.}\label{fig:example}\vspace*{-.4cm}
\end{figure}

We propose TCP/NC \cite{tcpnc,analysis} as such an alternate transport protocol. TCP/NC uses network coding, and modifies TCP's acknowledgment (ACK) scheme such that random erasures do not affect the transport layer's performance. TCP/NC receiver acknowledges \emph{degrees of freedom} instead of individual packets as shown in Figure \ref{fig:example}. Once enough degrees of freedoms are received at the receiver, the decoder solves the set of linear equations to decode the original data. TCP/NC may not be the only viable solution, and other transport protocols that can combat erasures may be used. We use TCP/NC for its effectiveness and simplicity.

TCP/NC allows a better use of the base stations installed, and can improve the goodput without any additional base stations. Improving the goodput with the same or a fewer number of base stations implies reduction in energy cost, operational expenses, capital expenses, and maintenance cost for the network provider. The results in this paper can also be understood as being able to serve more users or traffic growth with the same number of base stations. This may lead to significant cost savings, and may be of interest for further investigation.

\section{Model}\label{sec:setup}

Consider a network with $n$ users. We assume that these $n$ users are in an area such that a single base station can cover them as shown in Figure \ref{fig:towers}. If the users are far apart enough that a single base station cannot cover the area, then more base stations are necessary; however, we do not consider the problem of coverage.

The network provider's goal is to provide a \emph{fair} service to any user that wishes to start a transaction. Here, by fair, we mean that \emph{every user is expected to receive the same average throughput}, denoted as $r_t$ Mbps. The network provider wishes to have enough network resources, measured in number of base stations, so that any user that wishes to start a transaction is able to join the network immediately and achieve an average throughput of $r_t$ Mbps. We denote $r_g$ to be the goodput experienced by the user. Note that $r_g \leq r_t$.

We denote $n_{bs}$ to be the number of base stations needed to meet the network provider's goal. We assume that every base station can support at most $R_{\max}$ Mbps (in throughput) and at most $N_{\max}$ active users simultaneously. In this paper, we assume that $R_{\max} = 300$ Mbps and $N_{\max} = 200$.

A user is \emph{active} if the user is currently downloading a file; \emph{idle} otherwise. A user decides to initiate a transaction with probability $p$ at each time slot. Once a user decides to initiate a transaction, a file size of $f$ bits is chosen randomly according to a probability distribution $P_f$. We denote $\mu_f$ to be the expected file size, and the expected duration of the transaction to be $\Delta = \mu_f/r_g$ seconds. If the user is already active, then the new transaction is added to the user's queue. If the user has initiated $k$ transactions, the model of adding the jobs into the user's queue is equivalent to splitting the goodput $r_g$ to $k$ transactions (each transaction achieves a rate of $r_g/k$ Mbps).

We denote $p_p$ to be the probability of packet loss in the network, and $RTT$ to be the round-trip time. In a wireless, $p_p$ and $RTT$ may vary widely. For example, wireless connection over WiFi may have $RTT$ ranging from tens of milliseconds to hundreds of milliseconds with loss rates typically ranging from 0-10\%. In a more managed network (such as cellular networks), $RTT$ are typically higher than that of a WiFi network but lower in loss rates.

\section{Analysis of the Number of Base Stations}\label{sec:analysis}
We analyze $n_{bs}$ needed to support $n$ users given throughput $r_t$ and goodput $r_g$. We first analyze $P(\Delta, p)$, the probability that a user is active at any given point in time. Given $P(\Delta, p)$, we compute the expected number of active users at any given point in time and $n_{bs}$ needed to support these active users.

Consider a user $u$ at time $t$. There are many scenarios in which $u$ would be active at $t$. User $u$ may initiate a transaction at precisely time $t$ with probability $p$. Otherwise, $u$ is still in the middle of a transaction initiated previously.

To derive $P(\Delta, p)$, we use the Little's Law. For a stable system, the Little's Law states that the average number of jobs (or transactions in our case) in the user's queue is equal to the product of the arrival rate $p$ and the average transaction time $\Delta$. When $\Delta p \geq 1$, we expect the user's queue to have on average at least one transaction in the long run. This implies that the user is expected to be active at all times. When $\Delta p < 1$, we can interpret the result from Little's Law to represent the probability that a user is active. For example, if $\Delta p = 0.3$, the user's queue is expected to have 0.3 transactions at any given point in time. This can be understood as the user being active for 0.3 fraction of the time. Note that when the system is unstable, the long term average number of uncompleted jobs in the user's queue may grow unboundedly. 
In an unstable system, we assume that in the long term, a user is active with probability equal to one.

Therefore, we can state the following result for $P(\Delta, p)$.
\begin{equation}\label{eq:probability}
P(\Delta, p) = \min\lbrace1, \Delta p\rbrace = \min\left\lbrace1, \frac{\mu_f}{r_g}\cdot p\right\rbrace.
\end{equation}

Given $P(\Delta, p)$, the expected number of active users is $n P(\Delta, p)$. We can now characterize the expected number of base stations needed as
\begin{equation}\label{eq:basestations}
n_{bs} = nP(\Delta, p) \cdot \max{ \left\lbrace \frac{r_t}{R_{\max}}, \frac{1}{N_{\max}} \right\rbrace }.
\end{equation}
In Equation \eqref{eq:basestations}, $\max{ \lbrace \frac{r_t}{R_{\max}}, \frac{1}{N_{\max}} \rbrace }$ represents the amount of base stations' resources (the maximum load $R_{\max}$ or the amount of activity $N_{\max}$) each active user consumes. The value of $n_{bs}$ from Equation \eqref{eq:basestations} may be fractional, indicating that actually $\lceil n_{bs}\rceil$ base stations are needed.

Note the effect of $r_t$ and $r_g$. As shown in Equation \eqref{eq:basestations}, increasing $r_t$ incurs higher cost while increasing $r_g$ reduces the cost. Therefore, when a network provider dedicates resources to increase $r_t$, the goal of the network provider is to increase $r_g$ proportional to $r_t$.

\section{Best Case Scenario}\label{sec:bestcase}

In an ideal scenario, the user should see a goodput $r_g = r_t$. In this section, we analyze this best case scnario with $r= r_t = r_g$. Once we understand the optimal scenario, we then consider the behavior of TCP and TCP/NC in Section \ref{sec:tcp}.
\subsection{Analytical Results}\label{sec:analysisresults}

\begin{figure}[tbp]
\begin{center}
\subfloat[$\mu_f$ = 3.2 MB]{\includegraphics[width=.23\textwidth]{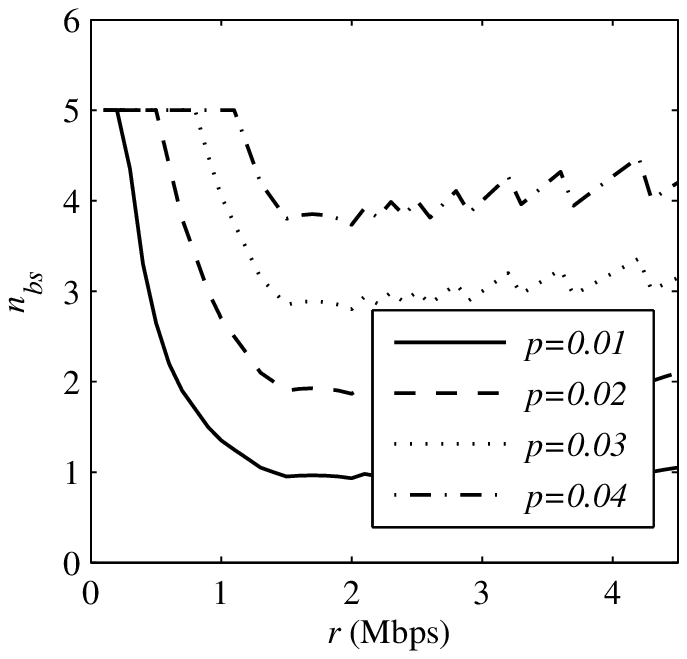}\label{fig:analysis_2}}
\subfloat[$\mu_f$ = 5.08 MB]{\includegraphics[width=.23\textwidth]{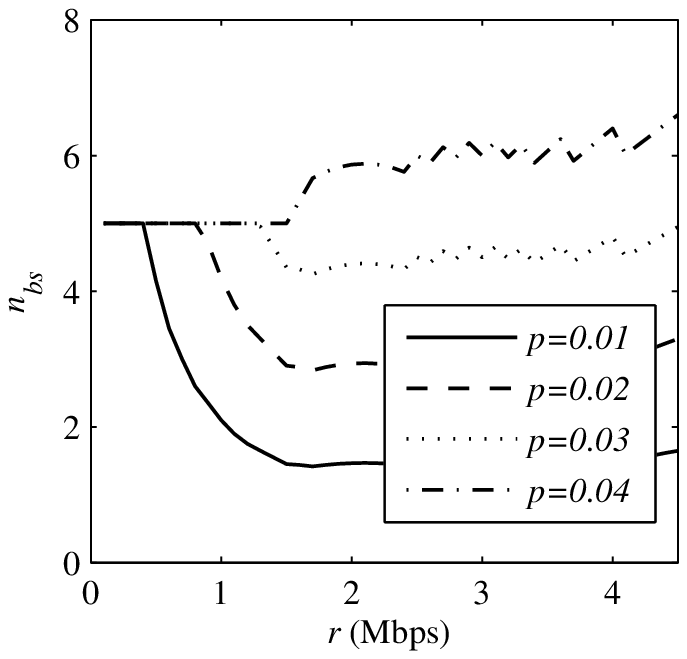}\label{fig:analysis_1}}
\end{center}\vspace*{-.2cm}
\caption{The values of $n_{bs}$ from Equation \eqref{eq:basestations} with $n=1000$ and varying $p$ and $r$.}\vspace*{-.4cm}
\end{figure}


In Figures \ref{fig:analysis_2} and \ref{fig:analysis_1}, we plot Equation \eqref{eq:basestations} with $\mu_f = 3.2$ MB and $\mu_f = 5.08$ MB for varying values of $p$. As $r$ increases, it does not necessarily lead to increase in $n_{bs}$. Higher $r$ results in users finishing their transactions faster, which in turn allows the resources dedicated to these users to be released to serve other requests or transactions. As a result, counter-intuitively, we may be able to maintain a higher $r$ with \emph{the same or a fewer} number of base stations than we would have needed for a lower $r$.
For example, in Figure \ref{fig:analysis_2}, when $r < 1$ Mbps, the rate of new requests exceeds the rate at which the requests are handled; resulting in an unstable system. As a result, most users are active all the time, and the system needs $\frac{n}{N_{\max}} = \frac{1000}{200} = 5$ base stations.

There are many cases where $n_{bs}$ is relatively constant regardless of $r$. For instance, consider $p=0.03$ in Figure \ref{fig:analysis_1}. The value of $n_{bs}$ is approximately 4-5 throughout. However, there is a significant difference in the way the resources are used. When $r$ is low, all users have slow connections; therefore, the base stations are fully occupied not in throughput but in the number of active users. On the other hand, when $r$ is high, the base stations are being used at full-capacity in terms of throughput. As a result, although the system requires the same number of base stations, users experience better quality of service and users' requests are completed quickly.

When $p$ and $r$ are high enough, it is necessary to increase $n_{bs}$. As demand exceeds the network capacity, it becomes necessary to add more infrastructure to meet the growth in demand. For example, consider $p = 0.04$ in Figure \ref{fig:analysis_1}. In this case, as $r$ increases $n_{bs}$ increases.


\subsection{Simulation Results}\label{sec:simulations}
\begin{figure}[tbp]
\begin{center}
\subfloat[$\mu_f$ = 3.2 MB]{\includegraphics[width=.23\textwidth]{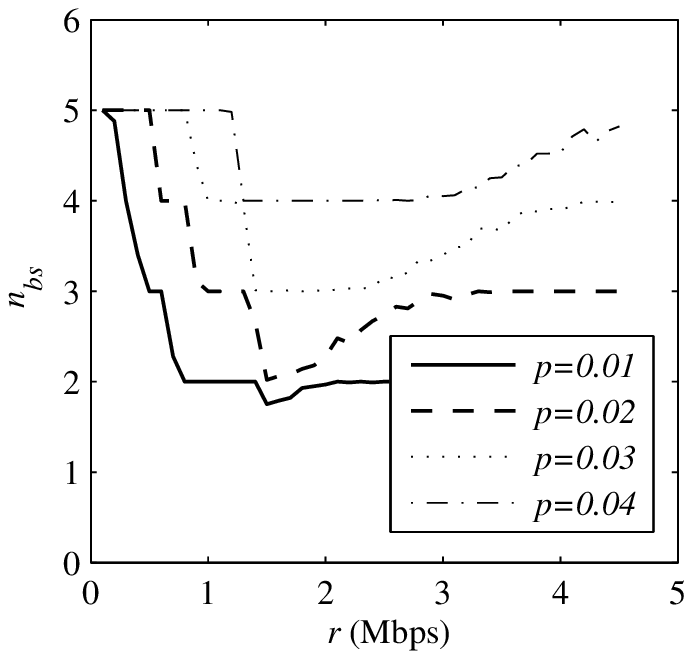}\label{fig:simulations_2}}
\subfloat[$\mu_f$ = 5.08 MB]{\includegraphics[width=.23\textwidth]{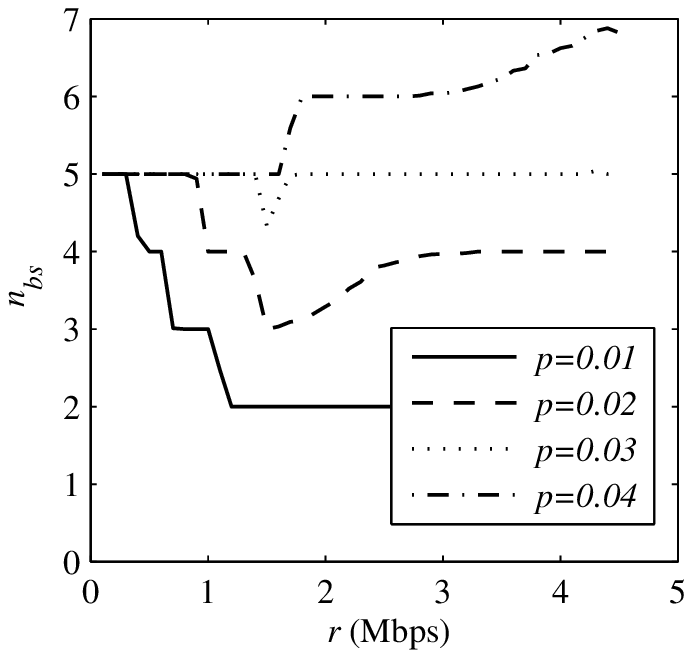}\label{fig:simulations_1}}
\end{center}\vspace*{-.2cm}
\caption{Average value of $n_{bs}$ over 100 iterations with $n=1000$ and varying $p$ and $r$.}\vspace*{-.4cm}
\end{figure}

We present MATLAB simulation results to verify our analysis results in Section \ref{sec:analysisresults}. We assume that at every 0.1 second, a user may start a new transaction with probability $\frac{p}{10}$. This was done to give a finer granularity in the simulations; the results from this setup is equivalent to having users start a new transaction with probability $p$ every second. We assume that there are $n=1000$ users. For each iteration, we simulate the network for 1000 seconds. Each plot is averaged over 100 iterations.

Once a user decides to start a transaction, a file size is chosen randomly in the following manner. We assume there are four types of files: $f_{doc}$ = 8KB (a document), $f_{image}$ = 1MB (an image), $f_{mp3}$ = 3 MB (a mp3 file), $f_{video}$ =  20 MB (a small video), and are chosen with probability $p_{doc}$, $p_{image}$, $p_{mp3}$, and $p_{video}$, respectively. In Figure \ref{fig:simulations_2}, we set $[p_{doc}, p_{image}, p_{mp3}, p_{video}] = [0.3, 0.3, 0.3, 0.1]$. This results in $\mu_f = 3.2$ MB as in Figure \ref{fig:analysis_2}. In Figure \ref{fig:simulations_1}, we set $[p_{doc}, p_{image}, p_{mp3}, p_{video}] = [0.26, 0.27, 0.27, 0.2]$, which gives $\mu_f = 5.08$ MB as in Figure \ref{fig:analysis_1}.

The simulation results show close concordance to our analysis. Note that the values in Figures \ref{fig:simulations_2} and \ref{fig:simulations_1} are slightly greater than that of Figures \ref{fig:analysis_2} and \ref{fig:analysis_1}. This is because, in the simulation, we round-up any fractional $n_{bs}$'s since the number of base stations needs to be integral.

\section{Analysis for TCP/NC and TCP}\label{sec:tcp}

We now study the effect of TCP and TCP/NC's behavior. We use the model and analysis from \cite{analysis} to model the relationship between $r_g$ and $p_p$ for TCP and TCP/NC. We denote $r_{g-nc}$ to be the goodput when using TCP/NC, and $r_{g-tcp}$ to be that for TCP. We set the maximum congestion window, $W_{\max}$, of TCP and TCP/NC to be 50 packets (with each packet being 1000 bytes long), and their initial window size to be 1. We consider $RTT$ = 100 ms and varying $p_p$ from 0\% to 5\%. We note that, given $r_t$ and $p_p$, $r_g \leq r_t (1-p_p)$ regardless of the protocol used.

In \cite{analysis,tcpnc}, TCP/NC has been shown to be robust against erasures; thus, allowing it to maintain a high throughput despite random losses. For example, if the network allows for 2 Mbps per user and there is 10\% loss rate, then the user should see approximately $2\cdot(1-0.1) = 1.8$ Mbps. Reference \cite{analysis} has shown, both analytically and with simulations, that TCP/NC indeed is able to achieve goodput close to 1.8 Mbps in such a scenario while TCP fails to do so.

\begin{figure}[tbp]
\begin{center}
\subfloat[$r_{g-nc}$]{\includegraphics[width=.23\textwidth]{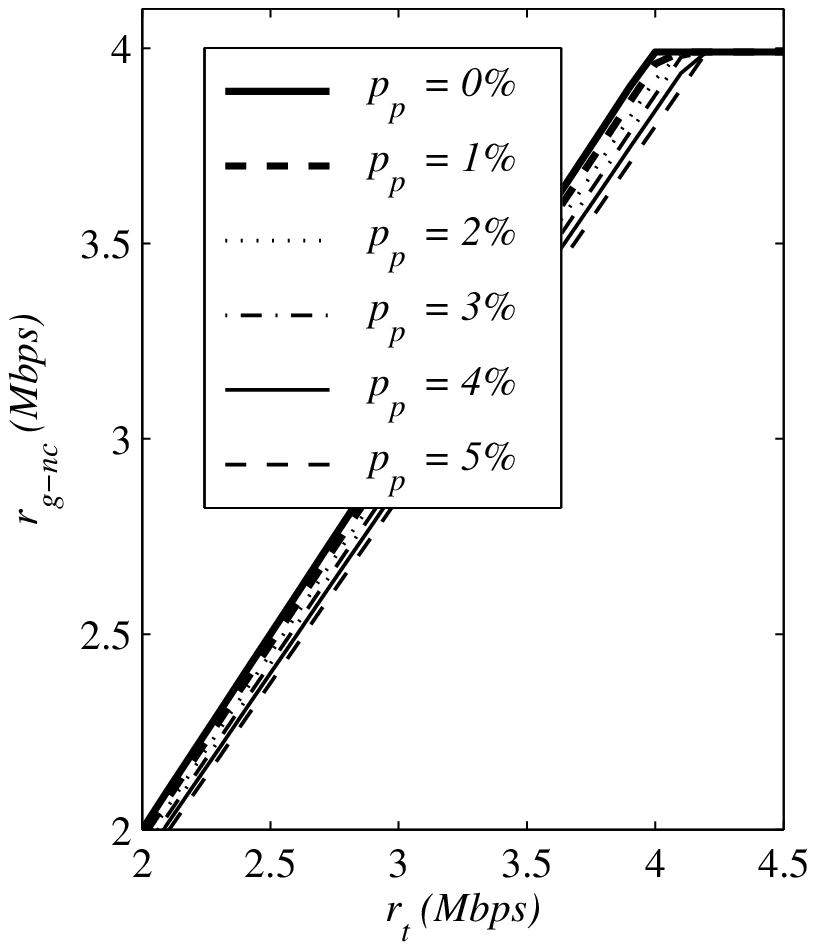}\label{fig:ncgoodput}}
\hspace*{.1cm}
\subfloat[$r_{g-tcp}$]{\includegraphics[width=.226\textwidth]{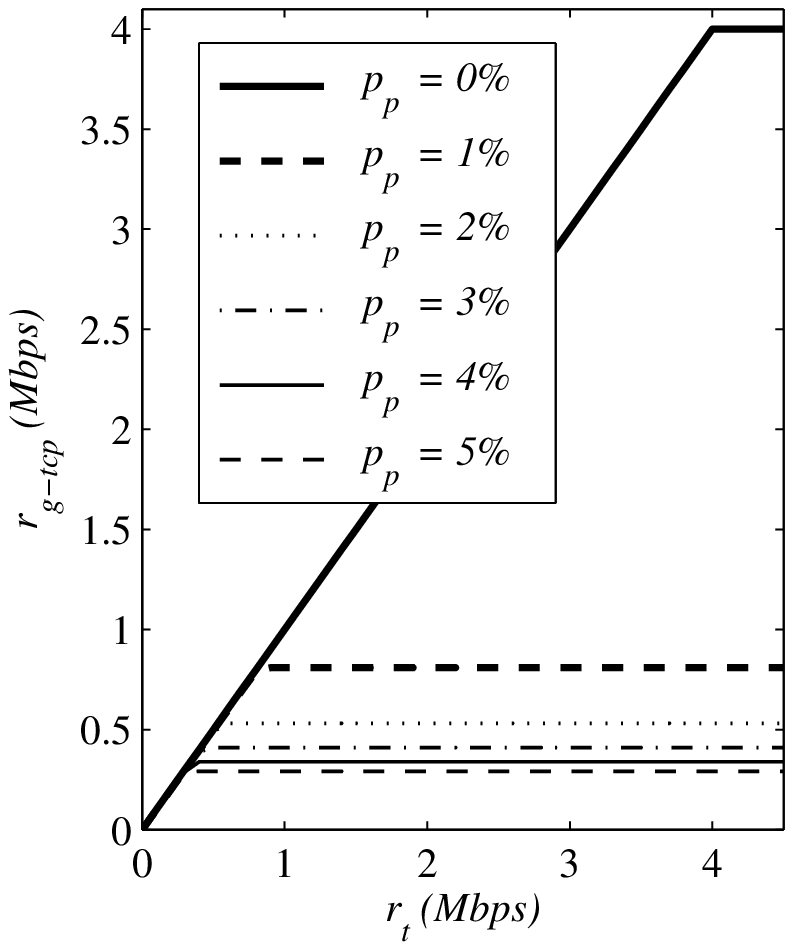}\label{fig:tcpgoodput}}
\end{center}\vspace*{-.2cm}
\caption{The value of $r_{g-nc}$ and $r_{g-tcp}$ against $r_t$ for varying values of $p_p$. We set $RTT$ = 100 ms.}\label{fig:goodput}\vspace*{-.3cm}
\end{figure}

\subsection{Behavior of $r_{g-nc}$ with varying $p_p$}\label{sec:ncgoodput}

\begin{figure*}[tbp]
\begin{center}
\subfloat[$p_p$ = 0\%]{\includegraphics[width=.22\textwidth]{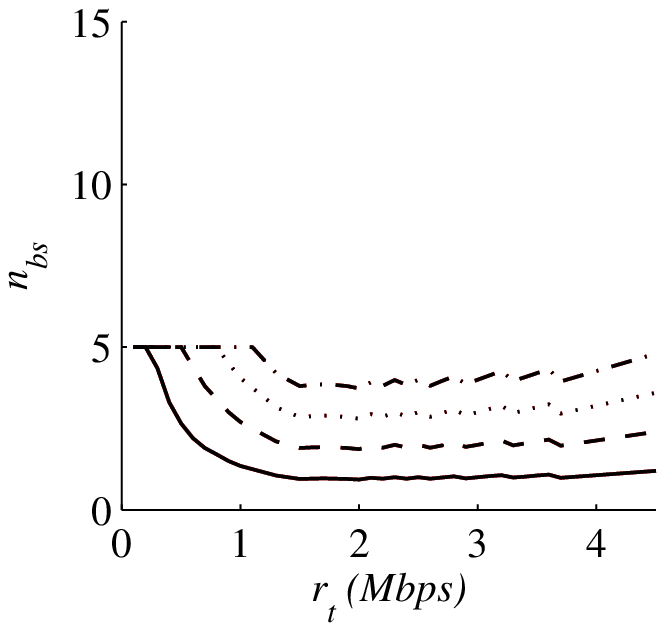}\label{fig:nbstcpnc100_320_0}}
\subfloat[$p_p$ = 1\%]{\includegraphics[width=.22\textwidth]{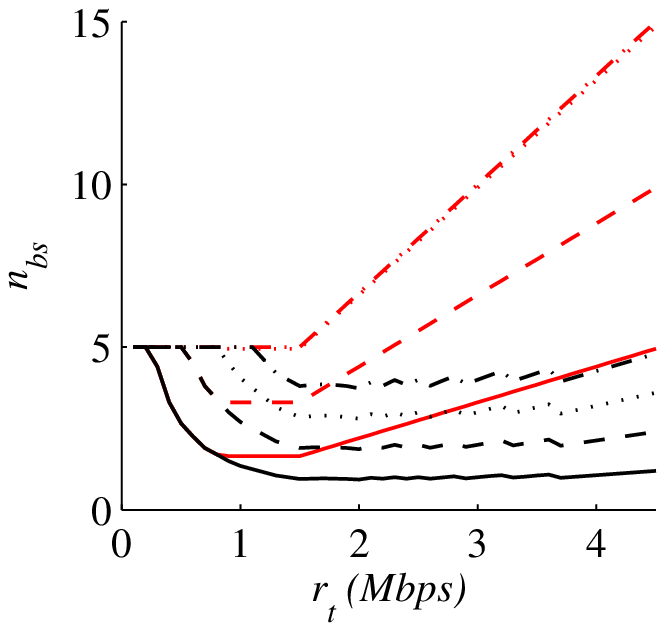}\label{fig:nbstcpnc100_320_1}}
\subfloat[$p_p$ = 2\%]{\includegraphics[width=.22\textwidth]{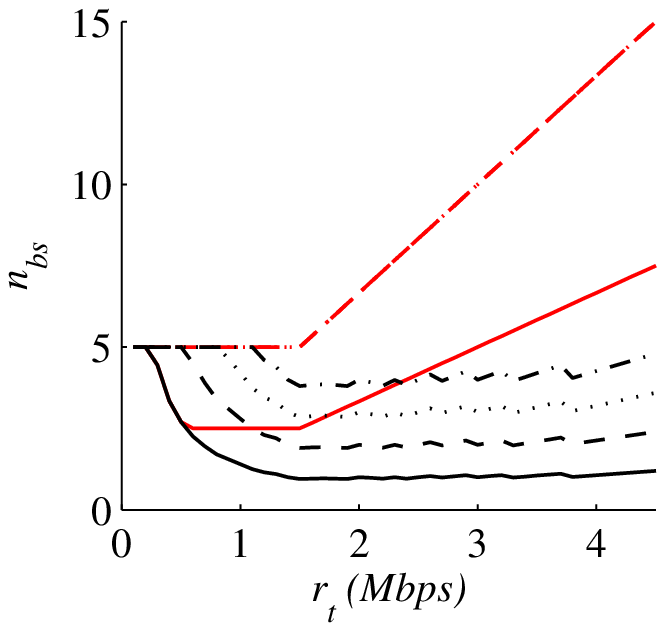}\label{fig:nbstcpnc100_320_2}}
\subfloat[$p_p$ = 5\%]{\includegraphics[width=.22\textwidth]{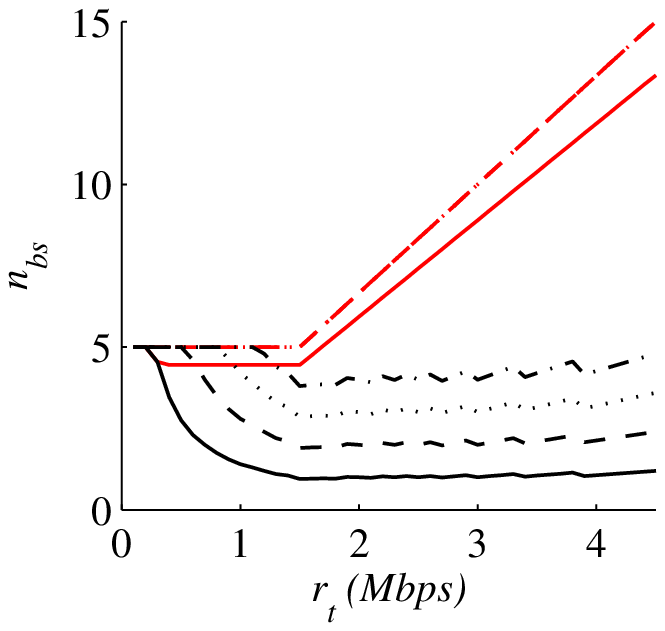}\label{fig:nbstcpnc100_320_5}}
\subfloat{\includegraphics[width=.12\textwidth]{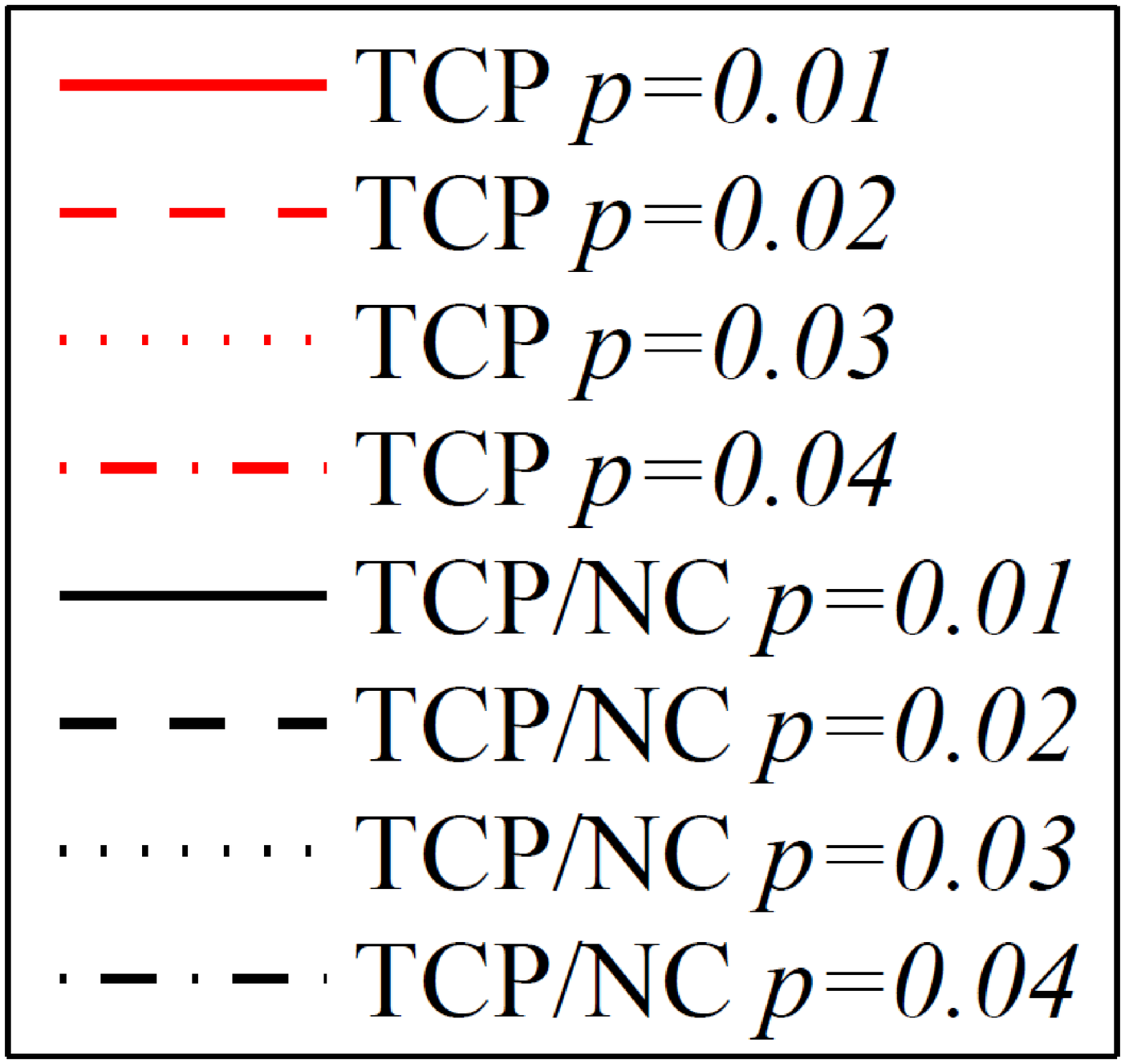}}
\end{center}\vspace*{-.2cm}
\caption{The value of $n_{bs}$ from Equation \eqref{eq:basestations} for TCP and TCP/NC with varying $p_p$ and $p$. Here, $RTT$ = 100 ms, $W_{\max} $ = 50, $n$ = 1000, and $\mu_f$ = 3.2 MB. In (a), $p_p = 0$ and both TCP and TCP/NC behaves the same; thus, the curves overlap. Note that this result is the same as that of Figure \ref{fig:analysis_2}. In (b), the value of $n_{bs}$ with TCP for $p = 0.03$ and 0.04 coincide (upper most red curve). In (c) and (d), the values of $n_{bs}$ with TCP for $p > 0.01$ overlap.}\label{fig:nbstcpnc100_320}\vspace*{-.3cm}
\end{figure*}

\begin{figure*}[tbp]
\begin{center}
\subfloat[$p_p$ = 0\%]{\includegraphics[width=.22\textwidth]{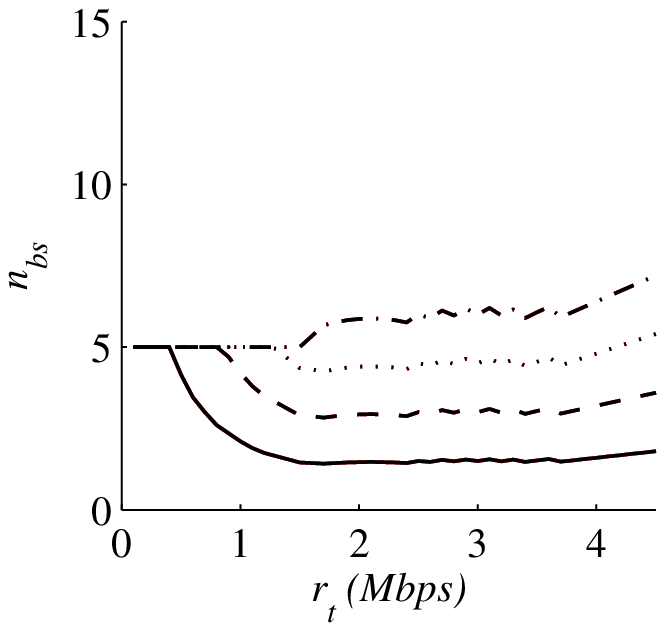}\label{fig:nbstcpnc100_580_0}}
\subfloat[$p_p$ = 1\%]{\includegraphics[width=.22\textwidth]{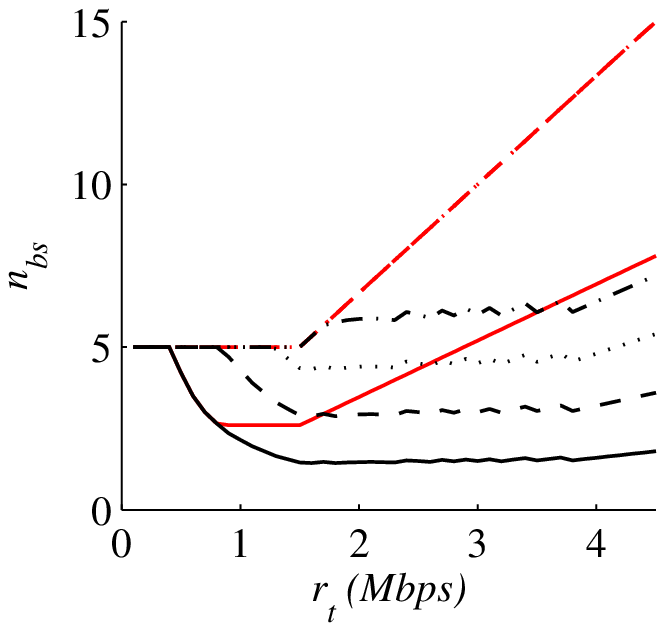}\label{fig:nbstcpnc100_580_1}}
\subfloat[$p_p$ = 2\%]{\includegraphics[width=.22\textwidth]{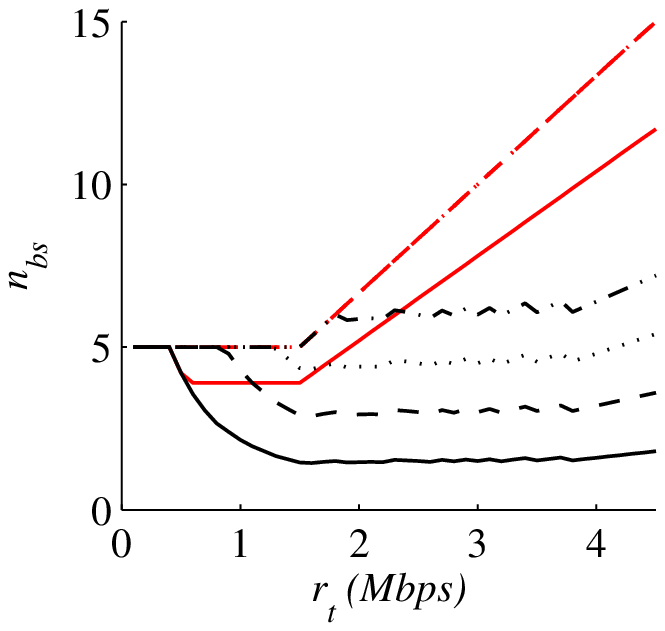}\label{fig:nbstcpnc100_580_2}}
\subfloat[$p_p$ = 3\%]{\includegraphics[width=.22\textwidth]{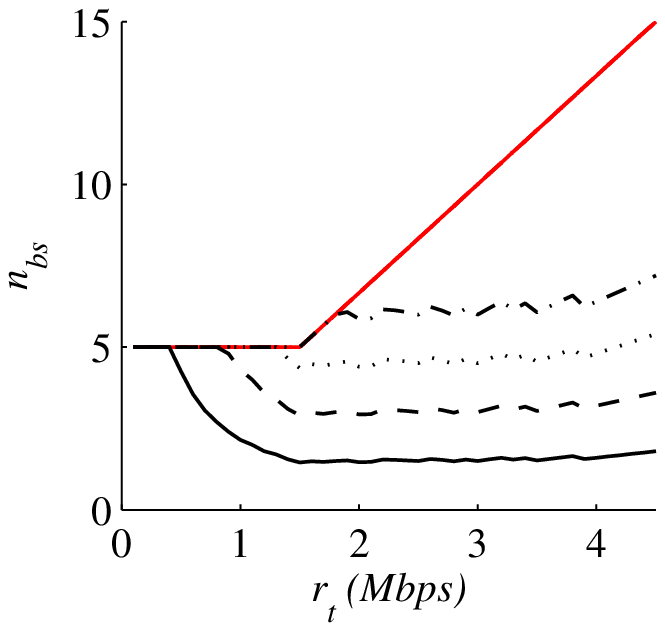}\label{fig:nbstcpnc100_580_3}}
\subfloat{\includegraphics[width=.12\textwidth]{legend}}
\end{center}\vspace*{-.2cm}
\caption{The value of $n_{bs}$ from Equation \eqref{eq:basestations} for TCP and TCP/NC with varying $p_p$ and $p$. Here, $RTT$ = 100 ms, $W_{\max} $ = 50, $n$ = 1000, and $\mu_f$ = 5.08 MB. In (a), the results for TCP and TCP/NC are the same. Note that this result is the same as that of Figure \ref{fig:analysis_1}. In (b) and (c), the value of $n_{bs}$ with TCP for $p> 0.01$ coincide (upper red curve). In (d), the values of $n_{bs}$ with TCP for any $p$ all overlap. We do not show results for $p_p$ = 4\% or 5\% as they are similar to that of (d).}\label{fig:nbstcpnc100_508}\vspace*{-.3cm}
\end{figure*}

Equation (20) from \cite{analysis} provides the goodput behavior of TCP/NC, which we provide below in Equation \eqref{eq:nc}.
\begin{equation}
\small
r_{g-nc} = \frac{1}{t SRTT} \left( tW_{\max} - \frac{(W_{\max} - 1)^2 + (W_{\max} - 1)}{2}\right),\label{eq:nc}
\end{equation}
where $SRTT$ is the effective RTT observed by TCP/NC and increases with $p_p$ and $t$ represents the duration of the connection (in number of RTTs). Equation \eqref{eq:nc} shows the effect of network coding. The goodput of TCP/NC decreases with $p_p$; however, the effect is indirect. As $p_p$ increases, the perceived RTT increases, which leads to TCP/NC reducing its rate.

Combining Equation \eqref{eq:nc} and 
$r_{g-nc} \leq r_t (1-p_p)$, we obtain the values of $r_{g-nc}$ for various $r_t$, $RTT$, and $p_p$. In Figure \ref{fig:ncgoodput}, the values of $r_{g-nc}$ plateaus once $r_t$ exceeds some value. This is caused by $W_{\max}$. Given $W_{\max}$ and $RTT$, TCP/NC and TCP both have a maximal goodput it can achieve. In the case with $RTT$ = 100 ms, the maximal goodput is approximately 4 Mbps. Note that regardless of $p_p$, all TCP/NC flows achieve the maximal achievable rate. This shows that TCP/NC can overcome effectively the erasures or errors in the network, and provide a goodput that closely matches the throughput $r_t$.

\subsection{Behavior of $r_{g-tcp}$ with varying $p_p$}\label{sec:tcpgoodput}

Equation (16) from \cite{analysis} provides the goodput behavior of TCP, which we provide below in Equation \eqref{eq:tcp}.
\begin{equation}
\small
r_{g-tcp} \approx \min\left(\frac{W_{\max}}{RTT}, \frac{1-p_p}{p_p}\frac{1}{RTT\left(\frac{5}{3} + \sqrt{ \frac{2}{3}\frac{1-p_p}{p_p}}\right)} \right).\label{eq:tcp}
\end{equation}
Note that unlike TCP/NC, TCP performance degrades proportionally to $\sqrt{\frac{1}{p}}$.

Combining Equation \eqref{eq:tcp} and $r_{g-tcp} \leq r_t (1-p_p)$, we obtain the values of $r_{g-tcp}$ for various $r_t$, $RTT$, and $p_p$ as shown in Figure \ref{fig:tcpgoodput}. As in Figure \ref{fig:ncgoodput}, the values of $r_{g-tcp}$ are also restricted by $W_{\max}$. However, TCP achieves this maximal goodput only when $p_p$ = 0\%. This is because, when there are losses in the network, TCP is unable to recover effectively from the erasures and fails to use the bandwidth dedicated to it. For $p_p > 0\%$, $r_{g-tcp}$ is not limited by $W_{\max}$ but by TCP's performance limitations in lossy wireless networks.


\subsection{The Number of Base Stations for TCP/NC and TCP}\label{sec:tcpncbs}

We use the values of $r_{g-nc}$ and $r_{g-tcp}$ from Sections \ref{sec:ncgoodput} and \ref{sec:tcpgoodput} to compare the number of base stations for TCP/NC and TCP using Equation \eqref{eq:basestations}. We assume that $SRTT = RTT$. In general, $SRTT$ is slightly larger than $RTT$.


Figures \ref{fig:nbstcpnc100_320} and \ref{fig:nbstcpnc100_508} show $n_{bs}$ predicted by Equation \eqref{eq:basestations} when $RTT$ = 100 ms. TCP suffers performance degradation as $p_p$ increases; thus, $n_{bs}$ increases rapidly with $p_p$. Note that increasing $r_t$ without being able to increase $r_g$ leads to inefficient use of the network, and this is clearly shown by the performance of TCP as $r_t$ increases with $p_p > 0\%$.

However, for TCP/NC, $n_{bs}$ does not increase significantly (if any at all) when $p_p$ increases. As discussed in Section \ref{sec:analysis}, TCP/NC is able to translate better $r_t$ into $r_{g-nc}$ despite $p_p > 0\%$, i.e. $r_t \approx r_{g-nc}$. As a result, this leads to a significant reduction in $n_{bs}$ for TCP/NC compared to TCP. Note that $n_{bs}$ for TCP/NC is approximately equal to the values of $n_{bs}$ in Section \ref{sec:analysis} regardless of the value of $p_p$. Since TCP/NC is resilient to losses, the behavior of $r_{g-nc}$ does not change as dramatically against $p_p$ as that of $r_{g-tcp}$ does. As a result, we observe $n_{bs}$ for TCP/NC to reflect closely the values of $n_{bs}$ seen in Section \ref{sec:analysis}, which is the best case with $r_t = r_g$.


We observe a similar behavior for other values of $RTT$ as we did for $RTT$ = 100 ms. The key effect of the value of $RTT$ in the maximal achievable goodput. For example, if $W_{\max}$ is limited to 50, the maximal achievable goodput is approximately 0.8 Mbps when $RTT$ = 500 ms, which is much less than the the 4 Mbps achievable with $RTT$ = 100 ms. As a result, for $RTT$ = 500 ms, neither $r_{g-nc}$ nor $r_{g-tcp}$ can benefit from the increase in $r_t$ beyond 0.8 Mbps. Despite this limitation, TCP/NC still performs better than TCP when losses occur. When demand exceeds the maximal achievable goodput, $n_{bs}$ increases for both TCP/NC and TCP in the same manner. We do not present the results for want of space.

\section{Conclusions}\label{sec:conclusions}

In wireless networks, the solution to higher demand is often to add more infrastructure. This is indeed necessary if all the base stations are at capacity (in terms of throughput). However, in many cases, the base stations are ``at capacity'' either because they are transmitting redundant data to recover from losses; or because they cannot effectively serve more than a few hundred active users. This may be 
costly as base stations are expensive to operate. One way to make sure that wireless networks are efficient is to ensure that, whenever base stations are added, they are added to effectively increase the goodput of the network.

We studied the number of base stations $n_{bs}$ needed to improve the goodput $r_g$ to the users. It may seem that higher $r_g$ necessarily increases $n_{bs}$. Indeed, if there are enough demand (i.e. $r_g$, $p$, or $\mu_f$ are high enough), we eventually need to increase $n_{bs}$. However, we show that this relationship is not necessarily true. When $r_g$ is low, each transaction takes more time to complete and each user stays in the system longer. This degrades the user experience and delays the release of network resources dedicated to the user. This is particularly important as the number of active users each base station can support is limited to the low hundreds. We observed that, given $r_t$, achieving low $r_g$ may lead to a significant increase in $n_{bs}$ and an ineffective use of the network resources; while achieving high $r_g$ may lead to reduction in $n_{bs}$. 


We showed that, in lossy networks, the goodput $r_g$ observed may not closely match the amount of resources dedicated to the user, e.g. $r_g \ll r_t$. This is due to the poor performance of TCP in lossy networks. To combat these harmful effects, network providers dedicate significant amount of resources, e.g. retransmissions and error corrections, to lower the loss rates. This, however, results in the base station transmitting at high throughput $r_t$ but little translating to goodput $r_g$. We showed that TCP/NC, which is more resilient to losses than TCP, may better translate $r_t$ to $r_g$. Therefore, TCP/NC may lead to a better use of the available network resources and reduce the number of base stations $n_{bs}$ needed to support users at a given $r_g$.

\bibliography{tcp_energy03}
\bibliographystyle{IEEEtran}

\end{document}